\documentclass[12pt]{article}
\pdfoutput=1

\usepackage{cite}

\usepackage[nointlimits,reqno]{amsmath}
\usepackage{amssymb}
\numberwithin{equation}{section} 

\usepackage{eucal}
\usepackage{bbm}
\usepackage{dsfont}

\usepackage{simplewick}
\usepackage{slashed}

\usepackage{graphicx}
\usepackage[pdftex,
 pdfproducer=TeXShop,
 pdfcreator=pdflatex]{hyperref}

\usepackage{xspace}
\usepackage{luty}

\begin{document}

\newcommand{\norm}[1]{|\!\gap|#1|\!\gap|}
\newcommand{\aavg}[1]{\avg{\!\avg{#1}\!}}
\newcommand{\1}{\mathds{1}} 

\begin{titlepage}

\title{Weyl versus Conformal Invariance\\
in Quantum Field Theory}

\author{Kara Farnsworth,\ \ Markus A. Luty,\ \ and\ \ Valentina Prilepina}

\address{Center for Quantum Mathematics and Physics (QMAP)\\
University of California, Davis, California 95616}

\begin{abstract}
We argue that conformal invariance in flat spacetime implies
Weyl invariance in a general curved background metric for all unitary theories
in spacetime dimensions $d \le 10$.
We also study possible curvature corrections to the Weyl transformations of operators,
and show that these are absent for operators of sufficiently low dimensionality
and spin.
We find possible `anomalous' Weyl transformations proportional to the 
Weyl (Cotton) tensor for $d > 3$ ($d = 3$).
The arguments are based on algebraic consistency conditions
similar to the Wess-Zumino consistency conditions that classify possible local 
anomalies.
The arguments can be straightforwardly extended to larger operator dimensions and higher $d$
with additional algebraic complexity.
\end{abstract}

\end{titlepage}

\section{Introduction}
Renormalization group (RG) fixed points in Poincar\'e invariant quantum field theory
are invariant under scale (dilatation) transformations $x^\mu \mapsto \la x^\mu$ 
by definition, but it is generally found that the spacetime symmetry is enhanced to
conformal symmetry, and even further to Weyl invariance when the theory is coupled
to a general background metric $g_{\mu\nu}$.
This enhancement has long been understood for theories derived from a scale-invariant
classical action \cite{Gross:1970tb, Callan:1970ze,Coleman:1970je, Iorio:1996ad},
but such theories are generally scale invariant at the quantum level only
for free field theories
or special theories (such as $\scr{N} = 4$ super Yang-Mills theory)
with exactly marginal interactions.
We will be interested in general IR fixed points where scale invariance
may be an accidental symmetry, and the fixed point is not necessarily described by a local scale invariant Lagrangian.
For example, the critical point of the 3D Ising model can be described
by the Landau-Ginzburg scalar field theory with tuned $\phi^2$ and $\phi^4$
terms in the Lagrangian.
This provides a UV Lagrangian description of the theory, but this Lagrangian
breaks scale invariance explicitly.
The IR fixed point is strongly coupled in terms of the scalar field,
and there is no known useful Lagrangian description of the fixed point.
Numerical studies of this theory indicate that it is conformally invariant 
\cite{Cosme:2015cxa, Kos:2016ysd};
our results show that any such theory is also Weyl invariant.

Conformal and Weyl invariance are closely related, and in fact are not always
clearly distinguished in the literature.
The response to an infinitesimal Weyl transformation 
$\de g_{\mu\nu} = 2\si g_{\mu\nu}$ is proportional to the trace of the
energy-momentum tensor $T = T^\mu{}_\mu$, so the vanishing of $T$ as
an operator statement in a general background metric implies Weyl invariance.
On the other hand, conformal invariance is the subgroup of Weyl transformations
that leaves the metric invariant up to a diffeomorphism.
The general enhancement of scale invariance in flat spacetime to 
conformal invariance in flat spacetime, and in turn to Weyl invariance in curved
spacetime has long been understood in $d = 2$ \cite{Polchinski:1987dy}.
In $d = 4$ there is a non-perturbative argument
\cite{Luty:2012ww,Dymarsky:2013pqa,Dymarsky:2014zja} that scale invariance
implies conformal invariance in flat spacetime, although it
has loopholes that in our view have not been satisfactorily closed 
\cite{Yonekura:2014tha}.
There is a much better understanding for $d = 4$ theories that can be viewed
as perturbations of a Weyl invariant fixed point, for example a free field theory.
For such fixed points, Weyl invariance is the only possible IR asymptotics
of the RG flow \cite{Jack:1990eb,Osborn:1991gm,Luty:2012ww,Fortin:2012hn,Baume:2014rla}.%
\footnote{If the IR fixed point contains an operator of dimension exactly
equal to 2, an improvement of the energy-momentum tensor is generally
required to obtain $T \equiv 0$.}
The perturbative arguments have been successfully extended to $d = 6$
\cite{Stergiou:2016uqq}, but attempts to generalize the non-perturbative
arguments have not been successful \cite{Elvang:2012st}.
There is a much better understanding for $d = 6$ theories with supersymmetry \cite{Cordova:2015fha}.
For a comprehensive review of the subject of scale versus conformal
symmetry, see \Ref{Nakayama:2013is}.

In this paper, we focus on the relation between conformal and Weyl invariance
in an arbitrary number of dimensions.
This question is interesting because Weyl transformations that are not conformal 
are commonly used in the literature, for example the Weyl transformation from flat spacetime
to the cylinder in $d > 2$ dimensions.
In this paper we will give a general non-perturbative argument that unitary 
conformally invariant quantum field theories are also
Weyl invariant.
Our argument holds for theories where the conformal generators are integrals
of local currents and for spacetime dimensions $d \le 10$.
Our argument starts with the fact that conformal invariance in flat spacetime
implies the vanishing of the trace of the energy-momentum operator $T$ in flat
spacetime, with the contact terms between $T$ and other operators generating
conformal transformations.
We then show that this implies that $T \equiv 0$ in curved space by systematically
classifying the possible corrections and imposing various algebraic consistency
conditions similar to the 
Wess-Zumino consistency conditions for Weyl anomalies.
The contact terms give the Weyl transformation of operators, and we show that
operators can be `covariantized' to have standard Weyl transformations,
at least for operators of sufficiently low dimension and spin.
It is straightforward to systematically extend the arguments in this paper
to higher spacetime dimensions and more general operators at the price of
additional algebraic complexity, but we do not attempt it here.

We identify possible consistent `anomalous' terms in the Weyl
transformation of operators, for example
\[
\eql{anomalousWeyltransintro}
\de_\si \scr{O} = -\De_{\scr{O}} \si \scr{O} + \si W^{\mu\nu\rho\si} W_{\mu\nu\rho\si} A,
\]
where $\scr{O}$ is a primary scalar operator with dimension $\De_{\scr{O}}$,
$A$ is a primary scalar operator (not the identity) with dimension $\De_{\scr{O}} - 4$, 
and $W_{\mu\nu\rho\si}$ is the Weyl tensor.
This is consistent because $W^{\mu\nu\rho\si} W_{\mu\nu\rho\si}$
transforms as a primary operator with dimension 4.
The existence of an operator $A$ with the required scaling dimension
is non-generic, and is allowed by unitarity constraints only
for $\De_{\scr{O}} \geq (d+6)/2$.
There are obvious generalizations of this to tensor operators made using the
Weyl tensor.
We note that these anomalous terms vanish for conformally flat metrics,
the case that is most commonly studied.
It is an open question whether there are any consistent anomalous
terms in the Weyl transformation for conformally flat metrics.

The existing literature on the question considered here is not extensive.
As already mentioned,
the question of whether conformal invariance implies Weyl invariance was
settled for $d = 2$ in \Ref{Polchinski:1987dy}.
Examples of non-unitary free field theories that are conformally invariant
but not Weyl invariant were discovered by mathematicians
\cite{GrahamConf,Eastwood:2002su,GoverConf}
and have been recently discussed in the physics literature 
\cite{Karananas:2015ioa,Brust:2016gjy}.
This work was largely inspired by \Ref{Karananas:2015ioa}, which we found
especially clear.
Other work on aspects of the relation between Weyl and conformal invariance includes
\Refs{Iorio:1996ad,Jackiw:2005su,Edery:2014nha,Karananas:2015eha}.

This paper is organized as follows.
In \S\ref{sec:ward} we state the problem precisely in terms of Ward identities
for conformal and Weyl invariance, and give a more detailed
outline of the argument.
In \S\ref{sec:flatspace}, we review some aspects of conformal invariance
in flat spacetime that we need for our argument.
In \S\ref{sec:curved} we give the main argument, showing that $T \equiv 0$
in a general curved spacetime, and hence the theory is Weyl invariant.
The details for $6 < d \le 10$ are given in an appendix.
We also constrain the possible Weyl transformations of operators
in this section.
In \S\ref{sec:examples} we discuss the non-unitary free field
theories that are conformally
invariant but not Weyl invariant, 
and use them to illustrate some of the steps of the general argument.
Our conclusions are given in \S\ref{sec:conclusions}.

\section{\label{sec:ward}
Conformal and Weyl Ward Identities}
Weyl invariance is defined for quantum field theories that can
be coupled to a background metric $g_{\mu\nu}$ in a diffeomorphism 
invariant way.%
\footnote{We expect that this holds in any theory that is sufficiently local in the UV.
It is known to fail in lattice models with sufficiently
non-local interactions, such as the long-range Ising model \cite{Picco:2012ak}.}
For such theories,
Weyl transformations are a local rescaling of the metric
combined with a transformation of the local operators.
For primary scalar operators $\scr{O}$, the transformation is
\[
\eql{Weyltransg}
\text{Weyl: }\ 
g_{\mu\nu}(x) &\mapsto \Om^2(x) g_{\mu\nu}(x),
\\
\eql{WeyltransPhi}
\scr{O}(x) &\mapsto \Om^{-\Delta_{\scr{O}}}(x) \scr{O}(x),
\]
where $\Om(x)$ is an arbitrary non-vanishing function of spacetime,
and $\Delta_{\scr{O}}$ is the dimension the operator $\scr{O}$.
Throughout this paper we focus on correlation functions of $\scr{O}$ for
simplicity.
Conformal transformations are special Weyl transformations such that the 
transformed metric is diffeomorphic to the original metric:
\[
\text{Conformal: }\ 
g_{\mu\nu}(x) &\mapsto \hat{\Om}^2(x) g_{\mu\nu}(x) = g'_{\mu\nu}(x)
\\
\scr{O}(x) &\mapsto \hat{\Om}^{-\Delta_{\scr{O}}}(x) \scr{O}(x),
\]
where $g'_{\mu\nu}$ is diffeomorphic to $g_{\mu\nu}$:
\[
g'_{\mu\nu}(x') = \frac{\d x^\rho}{\d x'^\mu} \frac{\d x^\si}{\d x'^\nu} g_{\rho\si}(x).
\]
The condition that conformal transformations are equivalent to diffeomorphisms 
places restrictions on the rescaling function $\hat{\Om}(x)$,
and a general metric will have no conformal symmetries.
We will consider conformally invariant theories in flat spacetime,
and we denote the flat spacetime metric by $\hat{g}_{\mu\nu}$.

It is clear from these definitions that Weyl invariance in a general
background metric implies conformal invariance in flat spacetime, but it is 
not at all obvious that the converse holds.
For $d > 2$ dimensions, the Euclidean conformal group is $SO(d+1, 1)$,
while the group of Weyl transformations is infinite-dimensional.
For $d = 2$ the conformal group is the infinite-dimensional Virasoro group, 
but the group of Weyl transformations is still larger.%
\footnote{The Weyl factor in a $d = 2$ conformal theory is a holomorphic
function, which implies that it satisfies the diffeomorphism invariant
constraint $\Box \Om = 0$.}

Weyl transformations relate correlation functions in different background metrics:
\[
\eql{WeylWard}
\begin{split}
\text{Weyl: }\ 
& \avg{\scr{O}(x_1) \cdots \scr{O}(x_n)}_{\Om^2 g_{\mu\nu}} 
\\[3pt]
&\qquad\quad\ {}
= \Om^{-\Delta_{\scr{O}}}(x_1) \cdots \Om^{-\Delta_{\scr{O}}}(x_n) 
\avg{\scr{O}(x_1) \cdots \scr{O}(x_n)}_{ g_{\mu\nu}}.
\end{split}
\]
On the other hand, conformal invariance in flat spacetime relates correlation
functions in the same metric:
%
%
\[
\eql{conformalWard}
\begin{split}
\text{Conformal: }\ 
&\avg{\scr{O}'(x_1) \cdots \scr{O}'(x_n)}_{\hat{g}_{\mu\nu}}
\\[3pt]
&\qquad\quad\ {}
= \hat{\Om}^{-\Delta_{\scr{O}}}(x_1) \cdots \hat{\Om}^{-\Delta_{\scr{O}}}(x_n)
\avg{\scr{O}(x_1) \cdots \scr{O}_n(x_n)}_{\hat{g}_{\mu\nu}},
\end{split}
\]
where $\scr{O}'$ is the image of $\scr{O}$ under a diffeomorphism:
\[
\scr{O}'(x') = \scr{O}(x).
\]
(Here we are neglecting possible conformal anomalies.
These will be included in the main argument below.)


We want to argue that \Eq{conformalWard} implies \Eq{WeylWard} 
for unitary quantum field theories.
It is useful to work with the infinitesimal form of the Ward identities,
which for the Weyl Ward identity is
\[
\eql{WeylWardinfinitesimal}
\!\!\!\!\!
\si(x) \avg{T(x) \scr{O}(x_1) \cdots \scr{O}(x_n)}_{g_{\mu\nu}}
= \sum_{i = 1}^n \de^d(x - x_i)
\avg{\scr{O}(x_1) \cdots \de_{\si} \scr{O}(x_i) \cdots \scr{O}(x_n)}_{g_{\mu\nu}},
\]
where
\[
\eql{WeylWardcontact}
\de_\si \scr{O} = -\Delta_{\scr{O}} \si \scr{O}
\]
is the infinitesimal operator transformation and $\si(x) = \ln \Omega(x)$.
Here $T$ is the trace of the energy-mo\-men\-tum tensor, defined in the standard way by
differentiation of the quantum effective action 
(generator of connected correlation functions) with respect to the background metric:
\[
\eql{energymomtensordefn}
\begin{split}
&\frac{\de}{\de g_{\mu\nu}(x)} \gap \frac{\de}{\de \rho(x_1)}
\left. \cdots \frac{\de}{\de \rho(x_n)} W_{\text{eff}}[g_{\mu\nu}, \rho] \right|_{\rho \, = \, 0}
\\
&\quad{}
= \left( -\frac{\sqrt{g(x)}}{2} \right) 
\left( -\sqrt{g(x_1)} \right) \cdots \left( -\sqrt{g(x_n)} \right) 
\avg{T^{\mu\nu}(x) \scr{O}(x_1) \cdots \scr{O}(x_n)}_{g_{\mu\nu}}.
\end{split}
\]
We will assume that the
quantum effective action $W_{\text{eff}}[g_{\mu\nu},\rho]$ 
is defined by a path integral
\[
e^{-W_{\text{eff}}[g_{\mu\nu}, \rho]} = \myint d[\Phi] \gap
e^{-\left( S[\Phi, g_{\mu\nu}] + \int\!\rho\gap \scr{O} \right)}.
\]
We do not assume that conformal symmetry is manifest at the level of the
path integral action $S$, so our arguments
apply to nontrivial conformal fixed points defined by a UV action that
is not conformally invariant, such as the critical point of the 3D Ising
model or the conformal window of QCD.
Our use of the path integral is limited to defining operators in terms
of sources, and operator redefinitions and contact terms
that are most conveniently expressed in terms of a path integral action.
These manipulations can be re-expressed in operator language independently
of the path integral, but we will not make this explicit.

To prove Weyl invariance, we must therefore prove two statements:
first, that $T \equiv 0$ up to contact terms,
and second, that the contact terms are given by \Eq{WeylWardcontact}.%
\footnote{We assume that the points $x_i$ are separated, so we do not have
to consider contact terms between the insertions of $\scr{O}$.}

We can now give a more detailed outline of our argument.
We first show that conformal invariance in flat spacetime implies 
$T \equiv 0$ in flat spacetime, possibly after improvement.
This is a standard result that is reviewed in the following section.
In curved spacetime there may be additional contributions to $T$
that depend on the spacetime curvature.
In \S\ref{sec:curved} we analyze these contributions, 
and show that they are associated with a symmetry of the effective action $W_{\text{eff}}$
that acts only on the sources that are used to define operators.
Algebraic closure of this symmetry and the unitarity inequalities 
on operator dimensions imply that $T \equiv 0$ in a general metric
for dimensions $d \le 10$.
The arguments can be straightforwardly
extended to higher dimensions at the price of additional
algebraic complexity.

Once we know that $T \equiv 0$ up to contact terms in a general metric, 
we can interpret the contact terms in
correlation functions of the form $\avg{T \scr{O\cdots O}}$ as
infinitesimal Weyl transformations of the correlation function $\avg{\scr{O \cdots O}}$.
These in turn are constrained by the fact that Weyl transformations commute.
Using this, we rule out additional terms in the Weyl transformation law for
operators of low dimension and spin, but find consistent anomalous Weyl transformations
in special cases, see \Eq{anomalousWeyltransintro}.
%

\section{\label{sec:flatspace}
Conformal Invariance in Flat Space}
In this section we review the standard result that 
in any conformally invariant theory 
we can define the energy-momentum tensor so that
$T \equiv 0$ in flat spacetime. We assume that in flat spacetime the conformal generators
$P_\mu$, $M_{\mu\nu}$, $D$, and $K_\mu$ are Hermitian operators
acting on the Hilbert space of the theory, and that these
operators are given by integrals of local currents:%
\footnote{Free $p$-form gauge theories with $d \ne 2(p+1)$ are examples
of scale invariant local quantum field theories
where the dilatation generator is not the integral of
a local current  \cite{ElShowk:2011gz,Dymarsky:2013pqa}.
These theories are not conformally invariant.
We are not aware of any
conformally invariant local quantum field theory in which the
conformal generators are not the integrals of currents.}
\[
Q = \myint d^{d-1} x\ggap J^0(x).
\]
Here the integral is over the surface $x^0 = \tau$, and we are using Cartesian
coordinates for flat space.%
\footnote{The arguments in this section can be straightforwardly
extended to general ``time'' surfaces in arbitrary coordinate systems.}
%
The conservation condition $\d_\mu J^\mu = 0$ ensures that the
integral is independent of $\tau$.
Assuming that the translation generators are given by
\[
P^\mu = \myint d^{d-1}x\ggap T^{\mu 0}
\]
and using the Euclidean Heisenberg equations of motion for the generators
\[
\frac{dQ}{d\tau} = [P^0, Q] + \frac{\d Q}{\d \tau},
\]
Wess \cite{WessCFT} showed that the current that gives the conformal generators
has the form
\[
\eql{conformalcurrent}
J^\mu =\xi^\nu T^\mu_\nu +( \d\cdot \xi)K^\mu  +\d_\nu (\d \cdot \xi ) L^{\mu\nu}.
\]
Here $\xi^\mu$ is the infinitesimal spacetime conformal transformation
parameter, given by 
\[
\xi^\mu = a^\mu + \om^\mu{}_\nu x^\nu + \la x^\mu + 2 (b \cdot x) x^\mu - x^2 b^\mu.
\]
The local operators $K^\mu$ and $L^{\mu\nu}$ have dimension $d-1$ and $d-2$
respectively.
Note that the antisymmetric part of $L^{\mu\nu}$ does not contribute
to $T$, so we assume that $L^{\mu\nu}$ is symmetric without loss of
generality.
Conservation of the current \Eq{conformalcurrent} then implies
\[\eql{TWess}
T  &= -\d_\mu K^\mu,
\\
K^\mu &= - \d_\nu L^{\nu\mu} .
\]
In $d = 2$, the infinite-dimensional Virasoro symmetry additionally requires that 
$L^{\mu\nu}$ be pure trace
\[
L_{\mu\nu} = \sfrac 12 L \de_{\mu\nu}
\quad (d = 2).
\]
%
%
The existence of the operator $L^{\mu\nu}$ (or $L$ in $d = 2$) implies that
we can redefine the energy-momentum tensor by adding the following `improvement'
terms to the action in the path integral in curved spacetime:
\[
\eql{improveflat}
\De S = \myint d^d x\gap \sqrt{g} \bigl[
\xi R L + \xi' R_{\mu\nu} L^{\mu\nu} \bigr],
\]
where $L = L^\mu{}_\mu$.
In flat spacetime these terms do not affect the dynamics of the theory,
but they change the definition of the energy-momentum tensor 
defined by functional differentiation with respect to the metric.
For $d = 2$, the second term  is redundant, and we set $\xi' = 0$.
The corrections to the energy-momentum tensor in flat spacetime can be obtained
from \Eq{improveflat} by expanding it to first order in metric perturbations 
about flat spacetime, so the metric dependence of $L^{\mu\nu}$
does not affect the correction to the energy-momentum tensor.
We obtain
\[
\De T^{\mu\nu} 
= -\bigl[ 2(d-1)\xi + \xi' \bigr] \partial^\mu \partial^\nu L    
-\sfrac{1}{2} (d-2) \xi' \left( \partial_\rho \partial^\mu L^{\nu\rho}+ \partial^\nu \partial_\rho L^{\mu\rho}\right) + O(R).
\]
By choosing
\[
\xi =  \frac{-1}{2(d-1)(d-2)}, 
\qquad \xi' = \frac{1}{(d-2)},
\qquad
(d \ge 3)
\]
or
\[
\xi = \frac{1}{2(d-1)}
\qquad
(d = 2),
\]
we obtain $T \equiv 0$ in flat spacetime.
In this way the vanishing of the trace of the (improved) energy-momentum tensor 
in flat spacetime follows from conformal invariance.

The above argument cannot be straightforwardly generalized to show that $T \equiv 0$ in 
a general background metric because such a metric generally has no conformal symmetries,
and these are a crucial ingredient in the argument.
Note also that the argument above does not assume unitarity of
the conformal field theory.
Unitarity will however be an essential ingredient in
our subsequent argument.

The operator relation $T \equiv 0$ is understood to hold up to
contact terms, and as discussed above, these contact terms give the transformation
of operators under Weyl and conformal transformations.
In the present case, once we know that $T \equiv 0$ up to contact terms, we 
can write the conformal generators as integrals of moments of the
energy-momentum tensor, for example
\[
K_\mu = 
\myint d^{d-1} x \gap \bigl( \de_{\mu\nu} x^2 - 2x_\mu x_\nu \bigr) T^{0\nu}.
\]
These obey the conformal algebra as a consequence of the tracelessness
condition $T \equiv 0$.
Using the conformal algebra and the assumption that $P_\mu$ acts by translation
on the fields, one can then derive the standard transformation properties of local operators
under conformal transformations \cite{Mack:1969rr}.
The conformal transformations of operators will be an important input to
the rest of our argument.

\section{\label{sec:curved}
Weyl Invariance in Curved Space}
We now consider the theory in a general curved background metric $g_{\mu\nu}$ and discuss
whether a quantum field theory that is conformally invariant in flat spacetime can
be shown to be Weyl invariant.

\subsection{\label{ssec:T0}
$T \equiv 0$ up to Contact Terms}
As discussed in \S\ref{sec:ward}, the first step in proving the Weyl Ward
identity \Eq{WeylWard} is to show that $T \equiv 0$ in curved spacetime, up to
contact terms.
Because $T(x)$ is a local operator that vanishes in flat spacetime, 
general covariance and locality require that it is 
proportional to at least one power of the Riemann curvature tensor
at $x$.
One possibility is that $T$ is proportional to powers of curvature tensors times
the identity operator, for example $T \propto c R \mathds{1}$
in $d = 2$.
This represents anomalous breaking of Weyl invariance, which will be discussed
in the following subsection.
For now we will focus on possible contributions to $T$ that are
proportional to nontrivial local operators,
for example $T = R X$, where  $R$ is the Ricci scalar, and $X$ is a scalar operator.
If such terms are present, then under a Weyl transformation the variation of the effective action $\de W_{\text{eff}}$ is non-local,
and there is no sense in which Weyl invariance is an approximate symmetry of the theory.
Note that in order to have $T = RX$, the operator
$X$ must have scaling dimension $d - 2$.
Scalar operators with such
special scaling dimensions are not generic in interacting conformal field theories.
Indeed, we will see that at every stage in our argument, the obstruction to
Weyl invariance involves the existence of operators with special scaling dimensions.
In a generic interacting theory, we do not expect to have operators with these special
dimensions.
However, our goal is to rule out these obstructions and obtain a completely
general result.

Let us consider the most general form for the operator correction to $T$.
The possible terms are limited by the unitarity constraints on the dimensions
of operators.
We first check that operator corrections to $T$ cannot involve non-scalar
primary operators, or their derivatives (descendant operators).
The reason is that any operator appearing in a curvature correction must have
dimension at most $d - 2$.
The unitarity constraints \cite{Mack:1975je,Minwalla:1997ka} 
exclude almost all higher
spin primary operators with dimension $\le d - 2$.
The only exception is an antisymmetric 
2-index tensor allowed for $d\geq 4$, which saturates the unitarity bound for $\De = d-2$,
but Lorentz invariance forbids any correction to $T$ in 
terms of such an operator.
Of course, descendants of higher-spin primary operators have even larger dimension,
and are therefore also excluded.
We conclude that in unitary theories the corrections to $T$ are proportional
to scalar primary operators or their descendants.
We can organize the possible terms in an expansion in powers of covariant
derivatives, where $R_{\mu\nu\rho\si} = O(\nabla^2)$:
\[
\eql{Tcurved}
\begin{split}
T &= R X 
\\
&\quad\ {}
+ R \Box Y_1
+ R^{\mu\nu} \nabla_\mu \nabla_\nu Y_2
+ \nabla^\mu R \nabla_\mu Y_3
+ \Box R Y_4
\\
&\quad\ {}
+ R^2 Y_5
+ R^{\mu\nu} R_{\mu\nu} Y_6
+ R^{\mu\nu\rho\si} R_{\mu\nu\rho\si} Y_7
\\
&\quad\ {}
+ O(\nabla^6).
\end{split}
\]
Here we used $\nabla^\mu R_{\mu\nu} = \frac 12 \nabla_\nu R$ to simplify the $O(\nabla^4)$
terms.
The operators $X$ and $Y_i$ in \Eq{Tcurved} are defined to be primary.
The operators $Y_i$ need not all be independent;
linear relations among them do not affect the argument below.
The unitarity bound on primary scalar operators is $(d-2)/2$, so the operator $X$
is allowed by unitarity for $d \ge 2$,
and the operators $Y_i$ are allowed for $d \ge 6$.
In general, we see that additional operators and higher powers of
derivatives are allowed for larger values of $d$.

Let us consider the case $d < 6$, in which case unitarity only allows $T = RX$.
For $d = 2$, this possibility can be excluded using the conservation of the
energy-momentum tensor \cite{Polchinski:1987dy}.
We give a general argument that does not depend on the special properties
of $d = 2$.
The idea is that the operator relation $T = RX$ 
reflects the existence of a nontrivial symmetry of the effective action $W_{\text{eff}}$.
The operators $T$ and $X$ are both defined by differentiation with
respect to sources, and the fact that this relation holds as an operator statement
tells us that these sources are not independent.
In other words, there is a redundancy in how the effective action
$W_{\text{eff}}$ depends on these sources,
which means that there is a symmetry that acts only on the sources.
We call this symmetry `Weyl redundancy.'
Symmetry transformations of this kind may be unfamiliar, 
so we illustrate them in various free field examples in \S\ref{sec:examples} below.

To define the operator $X$, we add a source term to the action
\[
\eql{Xsource}
\De S = \myint d^d x \gap \sqrt{g} \ggap \rho_X X.
\]
Then \Eq{Tcurved} implies that the quantum effective action $W_{\text{eff}}$ is invariant under
\[
\eql{sourcesymmetry}
\de_\si g_{\mu\nu} = 2 \si g_{\mu\nu},
\qquad
\de_\si \rho_X = \si R,
\]
where $\si$ is a general function of $x$.
Invariance under this transformation is 
what is required to reproduce $T = RX$, even though
the source term \Eq{Xsource} by itself is \emph{not} invariant.
Invariance of the effective action under \Eq{sourcesymmetry} is therefore
a very strong condition, and in fact can be easily ruled out.
The idea is that if \Eq{sourcesymmetry} is a symmetry of the effective action,
then the commutator of two such transformations is also a
symmetry.
Computing the commutators gives
\[
\eql{commutatorsigma12}
[\de_{\si_1}, \de_{\si_2}] g_{\mu\nu} = 0,
\qquad
[\de_{\si_1}, \de_{\si_2}] \rho_X = 2(d-1) (\si_1 \Box \si_2 - \si_2 \Box \si_1).
\]
For general $\si_1$ and $\si_2$ the function
$\si_1 \Box \si_2 - \si_2 \Box \si_1$
is an arbitrary function of $x$.
\Eq{commutatorsigma12} therefore implies
that $W_{\text{eff}}$ is invariant under $\rho_X \to \rho_X + \de\rho_X$ for an
arbitrary function $\de\rho_X$, with all other sources held fixed.
This in turn means that $W_{\text{eff}}$ is independent of $\rho_X$,
{\it i.e.\/}~the operator $X$ is trivial, proving that $T \equiv 0$ after all.

Note that the existence of the operator $X$ with dimension $d-2$ 
also allows us to add an `improvement'
term to the action
\[
\eql{Ximprove}
\De S = \myint d^d x \sqrt{g} \ggap \xi R X.
\]
However, this modifies $T$ in flat spacetime as well as curved spacetime
\[
\De T = \xi \bigl[ - 2(d-1) \Box X + (d-2) R X \bigr],
\]
and therefore plays no role in our argument.
A famous example of a conformal field theory with a primary operator $X$ with
dimension $d  - 2$
is free scalar field theory with $X = \frac 12 \phi^2$. 
An improvement term of the form \Eq{Ximprove} is required to make $T \equiv 0$
in flat spacetime, and then one finds that $T \equiv 0$ in an arbitrary curved
background.
In \S\ref{sec:examples} below this standard result
is rederived using the language of Weyl redundancy.

Let us now extend this argument to $d \ge 6$.
The case $d = 6$ is special because the operators $Y_i$ in \Eq{Tcurved} saturate 
the unitarity bound for scalar operators, and are therefore free scalar fields.
This means that each such operator generates a decoupled free scalar subsector
of the conformal field theory.
Each decoupled subsector has a separate conserved energy-momentum tensor,
and for each one we can use the arguments above.
The free fields $Y_i$ cannot appear in the energy-momentum tensor
for the interacting subsectors of the theory, so we conclude that
$T \equiv 0$ for interacting conformal field theories in $d = 6$.
Of course the free scalar subsectors are Weyl invariant
with suitable improvement of the energy-momentum tensor.

For $d > 6$ the argument becomes more complex.
There are more operators to consider (see \Eq{Tcurved}),
some of which can be improved away.
The generalization of the symmetry \Eq{sourcesymmetry} involves more operators
and sources, and the condition that $[\de_{\si_1}, \de_{\si_2}]$ is a symmetry is not
immediately sufficient to eliminate all possible corrections to $T$.
Nonetheless, we can use the fact that the metric can be chosen arbitrarily
to argue that all the corrections to $T$ vanish, at least for $d \le 10$.
The details of this argument are given in the appendix.
We will not attempt to extend this argument to higher values of $d$.
This is purely a matter of algebra, and is of limited interest since
we do not expect to have interacting conformal field theories
for such high dimensions in any case.

\subsection{Weyl Anomalies}
For even $d$, we can also have
curvature-dependent contributions to $T$ that are proportional
to the identity operator $\1$.
For example, in $d = 4$ the most general form allowed by scale invariance
and diffeomorphism invariance is 
\[
\eql{d4anomaly}
T = \bigl( c_1 R^2 + c_2 R^{\mu\nu} R_{\mu\nu} + c_3 R^{\mu\nu\rho\si} R_{\mu\nu\rho\si}
+ c_4 \Box R \bigr) \1.
\]
Because $T$ is the response of the theory to a Weyl transformation
$\de g_{\mu\nu} = 2\si g_{\mu\nu}$, \Eq{d4anomaly} is equivalent to
a local change in the effective action:
\[
\eql{conformalanomaly}
\de W_{\text{eff}} = -\myint d^4 x \gap \sqrt{g} \ggap \si 
\bigl( c_1 R^2 + c_2 R^{\mu\nu} R_{\mu\nu} + c_3 R^{\mu\nu\rho\si} R_{\mu\nu\rho\si}
+ c_4 \Box R \bigr)
\]
If Weyl invariance is broken only by a local $\de W_{\text{eff}}$, we say that the symmetry
has a Weyl anomaly \cite{Capper:1974ic, Deser:1976yx,Deser:1993yx}.
Despite the anomaly, the Weyl Ward identities still hold in a modified form, and Weyl invariance can in many ways still be regarded
as a good symmetry.
Weyl anomalies are necessarily present in even 
dimensions, for example they are nonzero even in free field theories.

The correction to $T$ above can be further constrained by imposing
the Wess-Zumino consistency conditions \cite{Wess:1971yu,Bardeen:1984pm, Cappelli:1988vw}.
We review it below
to highlight the similarities with the arguments above.
The first step is to note that we can cancel the term proportional to 
$\Box R$ by adding a local `improvement' term to the effective action
\[
\De W_{\text{eff}} = -\frac {c_4}{12} \myint d^4 x \gap \sqrt{g} \ggap  R^2.
\]
The $\Box R$ term in \Eq{conformalanomaly} can therefore be improved away,
and does not represent a
genuine anomaly.
The next step is to impose the constraint that Weyl transformations
commute, and therefore this must be reflected in $\de W_{\text{eff}}$.
To state the result, we change the basis of allowed curvature invariants
in \Eq{conformalanomaly} to
\[
\eql{conformalanomalybasis}
\de W_{\text{eff}} =- \myint d^4 x \gap \sqrt{g} \ggap \si 
\bigl(  a E_4 + b R^2 + c W^{\mu\nu\rho\si} W_{\mu\nu\rho\si} \bigr),
\]
where $E_4$ is the 4-dimensional Euler density.
One then finds
\[
[\de_{\si_1}, \de_{\si_2}] W_{\text{eff}} = -24 b \myint d^4 x \gap \sqrt{g} \ggap  (\si_1 \Box \si_2
- \si_2 \Box \si_1) R,
\]
so we must have $b = 0$, while the terms 
proportional to $c$ and $a$ in \Eq{conformalanomalybasis} are allowed.
We see that the arguments of the previous subsection are closely related to those
used to determine the most general form of the Weyl anomaly.

\subsection{\label{ssec:Weyltrans}
Contact Terms and Weyl Transformations of Operators}
The arguments up to now show that (at least for $d \le 10$)
$T \equiv 0$ in an arbitrary curved background metric,
but only up to contact terms.
As explained in the introduction, the contact terms in
the Weyl Ward identity \Eq{WeylWard} define the transformation of
local operators under Weyl transformations.
In this sense, we have already established Weyl invariance of the theory,
but we have not shown that operators transform in the canonical way.
In this section we analyze the structure of the contact terms, and show
that the possible Weyl transformations are highly constrained.
We are able to show that they have the canonical form except for a few
`anomalous' transformation laws that we are not able to exclude.
The main constraint comes from the fact that primary operators 
transform canonically under conformal transformations in flat space.
These transformations can be viewed as a special class of Weyl transformations.
Further algebraic consistency constraints come from the fact that Weyl 
transformations commute.

We now give some more detail about the connection between contact terms
and Weyl transformation of operators.
The most general contact terms in correlation functions with a single 
insertion of $T$ have the form
\[
\eql{almostWeylWard}
\begin{split}
\si(x) \avg{T(x) & \scr{O}(y_1) \cdots \scr{O}(y_n)}_{g_{\mu\nu}}
\\[2pt]
&
{}= \sum_{i = 1}^n \de^d(x - y_i) \avg{\scr{O}(y_1) \cdots 
\de_\si \scr{O}(y_i) \cdots \scr{O}(y_n)}_{g_{\mu\nu}}.
\end{split}
\]
This equation defines the local operator $\de_\si \scr{O}$, which
depends linearly on $\si$.
We consider the case where the $y_i$ in \Eq{almostWeylWard}
are separated points, so there are no
contact terms between the $\scr{O}$'s.
Because inserting $T$ is the response to a Weyl transformation, this equation
shows that the theory is Weyl invariant, with $\scr{O}$ transforming
under a Weyl transformations as $\scr{O} \mapsto \scr{O} + \de_\si \scr{O}$.%
\footnote{The connection between insertions of $T$ and Weyl transformations is
slightly more subtle at higher orders in $\si$, and will be discussed below.}
This is the sense in which we have already proven Weyl invariance, but note that 
we have not proven that the Weyl transformation of $\scr{O}$ is given by the 
standard formula $\de_\si\scr{O} = -\Delta_{\scr{O}} \si \scr{O}$.

In \Eq{almostWeylWard}, we allow $\de_\si \scr{O}$ to depend on derivatives of
$\si$.
That is, we allow terms such as $\de_\si \scr{O} = \Box \si B + \cdots$,
and we cannot cancel the $\si$ dependence on both sides of \Eq{almostWeylWard}.
The reason we must allow such terms because operators such as $\scr{O}$ and $T$ are really
distributions, and only smeared operators such as
\[
T[\si] = \myint d^d x\gap \sqrt{g(x)} \gap \si(x) T(x)
\]
are well-defined.
Specifically, a Weyl transformation is given by
\[
\eql{generalWeylWard}
\de_\si \avg{\scr{O}(y_1) \cdots \scr{O}(y_n)}_{g_{\mu\nu}}
&= \avg{T[\si] \scr{O}(y_1) \cdots \scr{O}(y_n)}_{g_{\mu\nu}}
\nn
&= \avg{\scr{O}(y_1) \cdots 
\de_\si \scr{O}(y_i) \cdots \scr{O}(y_n)}_{g_{\mu\nu}}.
\]

We will need to extend the connection between insertions of $T$ and the 
response to Weyl transformations beyond the linear order in $\si$.
It is then convenient to redefine $T$ to be the response to a Weyl transformation.
That is, we define
\[
\begin{split}
\frac{\de}{\de \si(x_1)} \cdots \frac{\de}{\de\si(x_m)} & \left.
\frac{\de}{\de\rho(y_1)} \cdots \frac{\de}{\de\rho(y_n)}
W_{\text{eff}} [e^{2\si} g_{\mu\nu}, \rho] \right|_{\substack{\si \, = \, 0 \\ \rho \, = \, 0}}
\\[4pt]
&= \left(-\sqrt{g(x_1)}\right) \cdots \left(-\sqrt{g(x_m)}\right) \left(-\sqrt{g(y_1)}\right) \cdots 
\left( -\sqrt{g(y_n)} \right)
\\
&\qquad\quad {}\times
\avg{T(x_1) \cdots T(x_m) \scr{O}(y_1) \cdots \scr{O}(y_n)}.
\end{split}
\]
This agrees with the previous definition \Eq{energymomtensordefn}
for correlation functions where all the points
$x_i$ and $y_i$ are separated.
That is, it differs from the previous definition only by contact
terms, so it does not affect the previous discussion.
For example, at quadratic order in $\si$ we now have\[
\begin{split}
\avg{\scr{O}(y_1) \cdots \scr{O}(y_n)}_{e^{2\si} g_{\mu\nu}}
&= \avg{\scr{O}(y_1) \cdots \scr{O}(y_n)}_{g_{\mu\nu}}
\\
&\qquad{}
+ \sum_{i = 1}^n \avg{\scr{O}(y_1) \cdots \de_\si \scr{O}(y_i) 
\cdots \scr{O}(y_n)}_{g_{\mu\nu}}
\\
&\qquad{}
+ \sum_{i < j}
\avg{\scr{O}(y_1) \cdots \de_\si \scr{O}(y_i) 
\cdots \de_\si \scr{O}(y_j)  \cdots \scr{O}(y_n)}_{g_{\mu\nu}}
\\
&\qquad\qquad{}
+ \sum_{i = 1}^n
\avg{\scr{O}(y_1) \cdots \de_\si \de_\si \scr{O}(y_i) 
\cdots \scr{O}(y_n)}_{g_{\mu\nu}}
\\
&\qquad{}
+ O(\si^3),
\end{split}
\]
where $\de_\si \de_\si \scr{O}$ is the contact term
between $T$ and $\de_\si \scr{O}$.
This tells us that $\de_\si \scr{O}$ fixes the Weyl variation of
the operator $\scr{O}$ to all orders in $\si$.

To proceed further, we use the conformal Ward identity \Eq{conformalWard} in flat spacetime.
Because a conformal transformation is the combination of a Weyl
transformation and diffeomorphism, subtracting the diffeomorphism 
contribution from the infinitesimal form of the Ward identity gives an equation
that is very similar to \Eq{almostWeylWard}:
\[
\eql{specialWeylWard}
\begin{split}
\hat{\si}(x) \avg{T(x) & \scr{O}(x_1) \cdots \scr{O}(x_n)}_{\hat{g}_{\mu\nu}}
\\[2pt]
&
{}= \sum_i \de^d(x - x_i) \avg{\scr{O}(x_1) \cdots 
\bigl[-\Delta_{\scr{O}} \hat{\si}(x_i) \scr{O}(x_i)\bigr] \cdots \scr{O}(x_n)}_{\hat{g}_{\mu\nu}}
+ \cdots
\end{split}
\]
The difference between this and the Weyl Ward identity is that this equation
holds only for a flat background metric $\hat{g}_{\mu\nu}$
and for a restricted class of Weyl parameters
\[
\hat{\si}(x) = \la + b \cdot x,
\]
where $\la$ is the parameter for dilatations,
$b_\mu$ is the parameter for special conformal transformations,
and $x^\mu$ are the standard Cartesian coordinates for flat Euclidean space.
We see that we must have $\de_\si \scr{O} \to -\Delta_{\scr{O}} \hat\si \scr{O}$ 
in the limit of flat spacetime and $\si \to \hat{\si}$.
We can then expand $\de_\si \scr{O}$ in a complete set of local operator expressions
linear in $\si$ that satisfy this condition.

To illustrate this, we consider the case where $\scr{O}$ is a relevant operator in $d \le 6$, 
in other words $\Delta_{\scr{O}} < d \le 6$.
In that case, the most general non-anomalous variation we can have is
\[
\eql{tildedeltaO}
\de_\si \scr{O} &= -\Delta_{\scr{O}} \si \scr{O}
+  \si R A  + \Box\si B.
\]
For example, a term of the form $\nabla_\mu \si V^\mu$ violates unitarity
for a primary vector operator $V^\mu$, while a term of the form
$\nabla_\mu \si \nabla^\mu C$ does not have the correct conformal transformation
in the flat space limit.
If we were to allow operators $\scr{O}$ with large scaling dimension, there would in
general be many additional terms in \Eq{tildedeltaO}.
Again, we note the appearance of operators with special dimensions, in this case
$\De_A, \De_B = \De_{\scr{O}} - 2$.
These operators are allowed by unitarity bounds for $\De_{\scr{O}} \geq (d+2)/2$.
We have neglected terms proportional to the identity operator,
which occur only for special values of $\Delta_{\scr{O}}$.
These are anomaly terms, and will be discussed below.

The unitarity bounds imply that $A$ and $B$ in \Eq{tildedeltaO} are conformal
primary operators (rather than descendants), and for $d \le 6$
do not allow any corrections to their transformation law analogous to \Eq{tildedeltaO}, so we have
\[
\de_\si A = -(\Delta_{\scr{O}} - 2) \si A,
\qquad
\de_\si B = -(\Delta_{\scr{O}} - 2) \si B.
\]
We can make a redefinition of the operator $\scr{O}$ by
\[
\eql{Ocovariantization}
\scr{O}' = \scr{O} + \frac{1}{2(d-1)} R B.
\]
The new operator transforms as
\[
\de_\si\scr{O}' = -\Delta_{\scr{O}} \si \scr{O}'
+ \si R A,
\]
so we do not have to consider the $\Box\si$ term in \Eq{tildedeltaO}.

Now the idea is that Weyl transformations commute, and so we must have
\[
[\de_{\si_1}, \de_{\si_2}] \scr{O}' = 0.
\]
Working out the commutator gives
\[
[\de_{\si_1}, \de_{\si_2}] \scr{O}' = -2(d-1)
(\si_2 \Box \si_1 - \si_1 \Box \si_2) A.
\]
Now $\si_2 \Box \si_1 - \si_1 \Box \si_2$ is an arbitrary function, 
so the operator $A$ must be trivial.
In this way, we have established that the operator $\scr{O}'$ has a standard
transformation under infinitesimal Weyl transformations.
We can regard the redefinition \Eq{Ocovariantization} 
as a `covariantization' of the operator $\scr{O}$
for Weyl transformations.

For larger values of $\De_{\scr{O}}$ there are consistent generalizations of 
the canonical transformation law, for example
\[
\eql{consistentOtransanomaly}
\de_\si \scr{O} = -\Delta_{\scr{O}} \si \scr{O} + \si W^{\mu\nu\rho\si} W_{\mu\nu\rho\si} A,
\]
where $A$ is a primary scalar operator
with $\De_A = \Delta_{\scr{O}} - 4$.
This is allowed by unitarity for $\De_{\scr{O}} \ge (d+6)/2$.
This is consistent because $W^{\mu\nu\rho\si} W_{\mu\nu\rho\si}$ 
has Weyl weight 4, and it cannot be eliminated by redefining $\scr{O}$.
This may therefore be viewed as an anomalous Weyl transformation for
the operator $\scr{O}$.
For $d = 3$, the Weyl tensor vanishes identically, but the Cotton tensor
\[
C_{\mu\nu\rho}= 
   \nabla_\rho \bigl(R_{\mu\nu} -  \sfrac{1}{4}g_{\mu\nu} R\bigr)
  - (\nu\leftrightarrow \rho)
\]
is Weyl invariant.
We can therefore have anomalous operator transformations of the form
\[
\eql{consistentOtransanomalyd3}
\de_\si \scr{O} = -\Delta_{\scr{O}} \si \scr{O} + \si C^{\mu\nu\rho} C_{\mu\nu\rho} A,
\]
where $\De_A = \De_{\scr{O}} - 6$.
Conformally flat metrics are characterized by the vanishing of the Weyl tensor
in $d > 3$, and the vanishing of the Cotton tensor in $d = 3$,
so these anomalies are absent in the conformally flat case.%
\footnote{%
A possible way to exclude 
\Eqs{consistentOtransanomaly} and \eq{consistentOtransanomalyd3}
is to use special metrics that are not conformally flat, 
but have nontrivial conformal isometries.
That is, the conformal Killing equation
$\nabla_\mu \xi_\nu + \nabla_\nu\xi_\mu = 2\si g_{\mu\nu}$
has solutions with $\si \ne 0$.
For each conformal Killing vector, we can define conformal generators acting
on fields using $T^{\mu\nu}$, as in flat spacetime.
If we can argue that these conformal transformations act on fields in the
standard way, we can exclude 
\Eqs{consistentOtransanomaly} and \eq{consistentOtransanomalyd3}.
We believe this may be a promising direction to explore.}
(In $d = 2$, all metrics are locally conformally flat.)

If $\De_{\scr{O}} = 2, 4, 6, \ldots$ we can have additional contributions to 
the transformation law proportional to the identity operator.
For example, for an operator of dimension 2 we must consider
\[
\de_\si \scr{O}_2 = -2 \si \scr{O}_2 + \left(c_1 \si R  + c_2\Box \si\right)\mathbbm{1}.
\]
We can redefine the operator
\[
\scr{O}'_2 &= \scr{O}_2 + \frac{1}{2(d-1)} c_2 R\mathbbm{1}
\]
so that 
\[
\de_\si \scr{O}'_2 = -2 \si \scr{O}'_2 +c_1 \si R \mathbbm{1}.
\]
This gives
\[
[\de_{\si_1}, \de_{\si_2}] \scr{O}'_2
\propto c_1 \bigl( \si_1 \Box \si_2 - \si_2 \Box \si_1 \bigr) \mathbbm{1},
\]
and therefore does not satisfy Weyl commutativity unless $c_1 = 0$.

For an operator of dimension 4, we can have the terms
\[
\eql{O4Weyltrans}
\begin{split}
\de_\si \scr{O}_4 &= -4 \si \scr{O}_4 + \si
\left( c_1 R^2 + c_2 R^{\mu\nu} R_{\mu\nu} + c_3 W^{\mu\nu\rho\si} W_{\mu\nu\rho\si} 
+ c_4 \Box R \right) \1
\\
&\qquad
+ \left( c_5 \nabla_\mu \si \nabla^\mu R + c_6 \Box \si R 
+ c_7 \nabla_\mu \nabla_\nu \si R^{\mu\nu}
+ c_8 \Box^2 \si \right) \1.
\end{split}
\]
We again can make a redefinition of the operator
\[
\scr{O}'_4 &= \scr{O}_4 + \left(
a_1 R^2 + a_2 R^{\mu\nu} R_{\mu\nu} + a_3 W^{\mu\nu\rho\si} W_{\mu\nu\rho\si} 
+ a_4 \Box R \right) \1
\]
to eliminate the $c_6, c_7, c_8$ terms in \Eq{O4Weyltrans}:
\[
\eql{O4pWeyltrans}
\begin{split}
\de_\si \scr{O}'_4 &= -4 \si \scr{O}'_4 + \si
\left( c_1 R^2 + c_2 R^{\mu\nu} R_{\mu\nu} + c_3 W^{\mu\nu\rho\si} W_{\mu\nu\rho\si} 
+ c_4 \Box R \right) \1
\\
&\qquad
+ c'_5 \nabla_\mu \si \nabla^\mu R \1,
\end{split}
\]
where
\[
c'_5 = c_5 + \frac{d-6}{2(d-1)} c_8.
\]
Commutativity of Weyl transformations then gives
\[
0 &= [\de_{\si_1}, \de_{\si_2}] \scr{O}'_4
\nn
&= \left[ 2(d-1) c_4 \si_1 \Box^2 \si_2 
- 2(d-1) c'_5 \nabla_\mu \Box \si_1 \nabla^\mu \si_2
- (1 \leftrightarrow 2) \right] \1 + O(R).
\]
Requiring Weyl commutativity in flat spacetime therefore implies that $c_4, c'_5 = 0$.
The curvature corrections then imply
\[
\begin{split}
0 &=\left[ 4(d-1) c_1 + 2 c_2  \right] R \si_1 \Box \si_2+ 2(d-2) c_2 R^{\mu\nu} \si_1 \nabla_\mu \nabla_\nu \si_2
- (1 \leftrightarrow 2).
\end{split}
\]
This must vanish for any $\si_1, \si_2$ in an arbitrary spacetime,
which gives $c_1, c_2 = 0$.
We find that the only possible anomaly has the form
\[
\de_\si \scr{O}_4 = -4 \si \scr{O}_4
+ c_3 \si W_{\mu\nu\rho\si}^2 \1.
\]
Such terms can be eliminated by the following argument.
The operator $\scr{O}_4$ can have a non-vanishing 1-point function, which by
locality and general covariance must take the form
\[
\avg{\scr{O}_4(x)}_{g_{\mu\nu}}
= \al R^2(x) + \be R_{\mu\nu}^2(x)
+ \ga W_{\mu\nu\rho\si}^2(x),
\]
for some coefficients $\al, \be, \ga$.
The infinitesimal form of the 
Weyl Ward identity \Eq{generalWeylWard} then tells us that
\[
\avg{\scr{O}_4}_{\Om^2 g_{\mu\nu}} - \avg{\scr{O}_4}_{g_{\mu\nu}}
= \avg{\de_\si \scr{O}_4}_{g_{\mu\nu}}
\]
or
\[
\de_\si \bigl[ \al R^2 + \be R_{\mu\nu}^2
+ \ga W_{\mu\nu\rho\si}^2 \bigr]
&= -4\si  \bigl[ \al R^2 + \be R_{\mu\nu}^2
+ \ga W_{\mu\nu\rho\si}^2 \bigr]
+ c_3 \si W_{\mu\nu\rho\si}^2.
\]
It is easily checked that this has no solution for a general metric
unless $c_3 = 0$.
Note that this argument uses the fact that the identity operator
necessarily has a non-vanishing 1-point function, and cannot be used to
rule out the anomalous transformations 
\Eqs{consistentOtransanomaly} and \eq{consistentOtransanomalyd3}
for $A \ne \mathbbm{1}$.

These arguments can be extended to higher dimension  
operators, operators with spin, and higher spacetime dimensions,
but it gets rapidly tedious.
To obtain a complete proof, one would try to proceed by induction starting
with the lowest dimensions and spins.
We will not attempt this here.
We have at least explicitly established 
that the Weyl transformation of relevant scalar operators for $d \le 6$
is the standard one.

\section{\label{sec:examples}
Examples}
In this section we consider the free field theories of 
\Refs{Karananas:2015ioa, Eastwood:2002su}, which can be used
to illustrate various aspects of the general arguments above.
These theories are defined by the action
\[
\eql{KMaction}
S = (-1)^{k+1} \myint d^d x \sqrt{g} \ggap \sfrac 12 \phi \Box^k \phi
\]
for $k = 1, 2, \ldots\,$.
The scalar field $\phi$ has dimension
\[
\De_\phi = \frac{d - 2k}{2},
\]
so these theories are non-unitary for $k > 1$.
\Ref{Karananas:2015ioa} showed that 
this theory is conformally invariant for all $k$ and $d$
in the sense that $T = \d_\mu \d_\nu L^{\mu\nu}$ in flat spacetime.
However, for special values of $d$ and $k$ the theory cannot be improved to be Weyl invariant
in curved spacetime:
\[
\begin{split}
k = 2 &: d = 2,
\\[-3pt]
k = 3 &: d = 2, 4,
\\[-3pt]
k = 4 &: d = 2, 4, 6,
\\[-6pt]
\vdots\ \ \ &
\end{split}
\]
In general, the theory cannot be coupled to gravity in a Weyl invariant way for 
all values of $k$ subject to the condition that $d$ is even and $d < 2k$. 
For these special theories the improvement terms required to obtain
$T \equiv 0$ in curved spacetime are divergent, so it is impossible
to make the theory Weyl invariant with a finite energy-momentum tensor.

All the special theories that are not Weyl invariant
have $\De_\phi < 0$, so these theories 
violate the unitarity bounds very badly.
For example, 2-point functions functions of $\phi$ grow with the separation.
We do not expect such theories to be relevant for physical statistical mechanics
systems.
In fact, as pointed out in \Ref{Brust:2016gjy},
the correlation functions of the theories
with $k > 1$ are not even scale invariant.
For example, the 2-point function satisfies
\[
\Box^k \avg{\phi(x)\phi(0)} = \de^d(x),
\]
which implies
\[
\begin{split}
\Box^{k-1} \avg{\phi(x)\phi(0)} &\propto \frac{1}{x^2},
\\
\Box^{k-2} \avg{\phi(x)\phi(0)} &\propto \ln x^2,
\\[3pt]
\Box^{k-3} \avg{\phi(x)\phi(0)} &\propto x^2 \ln x^2 - 3 x^2,
\end{split}
\]
{\it etc\/}.
These logarithms represent genuine non-local breaking of scale invariance.
For example, for $k = 2$ we have $\avg{\phi(x)\phi(0)} \propto \ln x^2$ and 
the effective action contains terms
\[
W_{\text{eff}} \sim \myint d^d x \gap d^d y \gap \rho_\phi(x) \rho_\phi(y) \ln [(x-y)^2] + \cdots,
\]
where $\rho_\phi$ is the source for $\phi$.
Under a scale transformation, we get non-local terms
\[
\de W_{\text{eff}} \sim \myint d^d x \gap d^d y \gap \rho_\phi(x) \rho_\phi(y) + \cdots.
\]
The correlation functions of this theory are clearly 
not scale invariant in any meaningful sense.%
\footnote{%
\Ref{Brust:2016gjy} defines a conformal theory algebraically using
$\avg{\phi(x)\phi(0)} \equiv (x^2)^{k - d/2}$ and 
defining correlation functions by Wick contraction.
We cannot take this approach for our purposes, because this theory has no 
local definition, and therefore cannot be
coupled to a metric.}

Although the $k > 1$ theories defined by \Eq{KMaction}
are not scale invariant as quantum theories,
the action is conformally invariant.
We can then ask whether the action can also be made Weyl invariant
by adding improvement terms.
Under a Weyl transformation, we transform both the metric 
and the fields $\phi$,
and the condition for Weyl invariance is
\[
0 = \frac{\de S}{\de \si(x)} = -T(x) + \De_\phi \phi(x) \gap \scr{E}(x) .
\]
where $\scr{E} = \de S / \de \phi$ is the equation of motion operator
and $\De_\phi$ is the Weyl weight of $\phi$.
On solutions to the classical equations of motion, the condition
of Weyl invariance is therefore equivalent to $T \equiv 0$, just as
for quantum theories.
We can therefore use the theories defined by \Eq{KMaction} as examples
of conformal invariance without Weyl invariance
in the classical limit, and use
them to illustrate some aspects of our general
argument.

\subsection{Weyl Redundancy}
In \S\ref{sec:curved} we argue that operator corrections to $T$ 
reflect the existence of a symmetry that acts only on sources,
which we call Weyl redundancy.
Such a symmetry was ruled out for unitary theories.
We now show that this symmetry does exist in the non-unitary theories defined by 
\Eq{KMaction}, explaining how they evade our argument. 

We start with the case $k = 1$, the usual free scalar.
We write the action for this theory as
\[
S = \myint d^d x \gap \sqrt{g} \left[
\sfrac 12 Z g^{\mu\nu} \d_\mu \phi \d_\nu \phi
+ \sfrac 12 m^2 \phi^2 + \sfrac 12 \xi R \phi^2 \right],
\]
where we have included an arbitrary improvement term as well as a mass term.
We consider $g_{\mu\nu}$, $Z$, $m^2$, and $\xi$ as spacetime dependent
background sources, although we are interested in the theory with $Z = 1$, $m^2 = 0$.
With these source terms, the action is invariant under the symmetry transformation
\[
\eql{weylredundk2}
\begin{split}
\de g_{\mu\nu} &= 2 \si g_{\mu\nu},
\\
\de Z &= -(d-2) \si Z,
\\
\de m^2 &= -d \si m^2 + 2 (d-1) \xi \Box \si,
\\
\de \xi &= -(d-2)\si \xi,
\\
\de \phi &= 0.
\end{split}
\]
Note that the fields do not transform in \Eq{weylredundk2}, 
so this is a redundancy among the sources.
The second term in the transformation of $m^2$ comes from the fact that
$\de R \supset -2(d-1)\Box \si$.
This symmetry implies the operator relation
\[
0 = \left. \frac{\de S}{\de \si} \right|_{\substack{{\,\,\,\, Z \, = \, 1} \\[2pt] m^2 \, = \, 0}}
 = -T - \frac{d-2}{2}  (\d\phi)^2
+ (d-1) \xi \Box\phi^2 - \frac{d-2}{2} \xi R \phi^2.
\]
Using the equation of motion
$\Box\phi = \xi R \phi$
gives
$(\d\phi)^2 = \sfrac 12 \Box\phi^2 - \xi R \phi^2$,
which implies
\[
T = \left[ -\frac{d - 2}{4} +  (d-1) \xi \right] \Box \phi^2.
\]
This vanishes with the choice $\xi  = \frac{(d-2)}{4(d-1)}$.
Note that the terms involving $R$ have canceled, reflecting the fact that the
same improvement can make  $T \equiv 0$ both in flat and curved spacetime.
This illustrates Weyl redundancy, and shows how it can be used to compute $T$.

A less trivial example is the $k = 2$ theory.
The presence of higher derivatives in the action means that the action depends
on the gravitational connection, and we do not obtain a simple
scaling symmetry of the form \Eq{weylredundk2}.
We can however make such a symmetry again manifest by
rewriting the action in terms of an auxiliary field $F$ so that it contains
only first derivatives of fields:
\[
\eql{newSk2}
S = \myint d^d x \gap \sqrt{g} \ggap \bigl[\sfrac 12 F^2 - F \Box \phi  \bigr].
\]
The equation of motion for $F$ is $F = \Box \phi$, 
and substituting this back into the action gives the original action \Eq{KMaction}.
We can now integrate by parts in \Eq{newSk2} to write
\[
S = \myint d^d x \gap \sqrt{g} \bigl[ \sfrac 12 Z F^2 + Z' g^{\mu\nu} \d_\mu F \d_\nu \phi  \bigr].
\]
This action is invariant under
\[
\begin{split}
\de g_{\mu\nu} &= 2 \si g_{\mu\nu},
\\
\de Z &= -d \si Z,
\\
\de Z' &= -(d-2) \si Z'
\end{split}
\]
with $\de\phi, \de F = 0$.
This symmetry implies the operator relation
\[
\eql{Tk2model1}
0 = \frac{\de S}{\de \si} 
= -T - \frac{d}{2} (\Box \phi)^2 - (d-2) \d\phi \d(\Box\phi).
\]
Using the equation of motion $\Box^2 \phi = 0$, we have
\[
\Box(\phi \Box \phi) &= (\Box\phi)^2 + 2 \d\phi \d(\Box\phi),
\\
\Box^2 (\phi^2) &= 2 (\Box\phi)^2 + 8 \d\phi \d(\Box\phi) + 4 (\d_\mu \d_\nu \phi)^2,
\\
\d_\mu \d_\nu (\phi \d_\mu \d_\nu \phi) &=
2 \d\phi \d (\Box\phi) + (\d_\mu \d_\nu \phi)^2,
\]
which we can use to write \Eq{Tk2model1} as
\[
T = \d_\mu \d_\nu \left(2\phi \partial^\mu \partial^\nu \phi -  \eta^{\mu\nu} (\partial \phi)^2 - \frac{d}{2} \eta^{\mu\nu} \phi\Box \phi  \right)
\]
in agreement with \Eq{TWess}.  Already we can see that this theory cannot be improved to be invariant under the full Virasoro algebra in $d = 2$. Note also that in both of these examples, the symmetry is Abelian ($[\de_{\si_1}, \de_{\si_2}] = 0$),
so the existence of the symmetry in these cases does not contradict the argument above.

We can extend these results to include
improvement terms to our action as well as additional source terms:
\[
S = \myint d^d x \gap \sqrt{g} \bigl[ \sfrac 12 Z F^2 + Z' g^{\mu\nu} \nabla_\mu F \nabla_\nu \phi + \sfrac 12 m^2 \phi^2 + \sfrac{1}{2} \kappa^{\mu\nu}\nabla_\mu \phi \nabla_\nu \phi \nn
\qquad{} +c_1 R \nabla_\mu\phi \nabla^\mu\phi + c_2 \Box R \phi^2 +c_3 R^{\mu\nu} \nabla_\mu \phi \nabla_\nu \phi\\
\qquad{} + c_4 R^2\phi^2 + c_5 R_{\mu\nu}^2\phi^2 + c_6 W_{\mu\nu\rho\sigma} ^2\phi^2 \bigr]\nonumber.
\]
This action is invariant under the more complicated transformation
\[
\begin{split}
\de g_{\mu\nu} &= 2 \si g_{\mu\nu},
\\
\de Z &= -d \si Z,
\\
\de Z' &= -(d-2) \si Z',
\\
\delta m^2 &= -d\sigma m^2+4(d-1)c_2  \nabla^4 \sigma-2(d-4)c_2\sigma \Box R-2(d-6)c_2\nabla_\mu \sigma \nabla^\mu R\\
& \qquad{}+4(d-2) c_5 R^{\mu\nu}\nabla_\mu \nabla_\nu \sigma   +4\left[c_2+2(d-1)c_4 +c_5\right]R \Box \sigma \\
& \qquad{}-2(d-4) \sigma c_4 R^2  - 2(d-4) \sigma c_5 R_{\mu\nu}^2- 2(d-4) \sigma c_6W_{\mu\nu\rho\sigma}^2,\\
\delta \kappa_{\mu\nu} &= -(d-4) \sigma \kappa_{\mu\nu}+2(d-2)c_3\nabla_\mu \nabla_\nu\sigma +2\left[2(d-1)c_1+c_3\right]g_{\mu\nu} \Box\sigma \\ 
& \qquad{}-2(d-4)c_3\sigma R_{\mu\nu}-2(d-4)c_1\sigma g_{\mu\nu} R.
\end{split}
\]
The choice
\[
\begin{split}
c_1&= \frac{-2d-(d-4)(d-2)}{4(d-1)(d-2)}\\
c_2&= \frac{(d-4)}{8(d-1)}\\
c_3 &=\frac{2}{(d-2)}
\end{split}
\]
guarantees that $T \equiv 0 $ in flat space, and gives 
\[
\begin{split}
T &= \left[ -\frac{(d-4)}{(d-2)}+2(d-2)c_5\right] \nabla_\mu \nabla_\nu(G^{\mu\nu} \phi^2)
\\
&\qquad\quad{}
+\left[\frac{(d-4)^2}{8(d-1)}+4(d-1)c_4 +dc_5\right]\nabla^2 (R \phi^2)
\end{split}
\]
in curved space once we use the improved equations of motion. We see that in this example we can choose $c_4$ and $c_5$ so that $T \equiv 0$ in curved space as long as $d \neq 2$. 
This confirms the results of \Refs{Karananas:2015ioa, Eastwood:2002su}
for the $k = 2$ theories, which are Weyl invariant unless $d = 2$.
It also illustrates the utility of Weyl redundancy in calculations.


\section{\label{sec:conclusions}
Conclusions}
We have given a general argument that conformal invariance in flat spacetime
implies Weyl invariance in curved spacetime in local unitary quantum field theories.
Conformal transformations are the subgroup of Weyl transformations that leave the
metric invariant (up to a diffeomorphism), 
so a failure of Weyl invariance arises from corrections that are non-vanishing for
curved backgrounds and/or general scale factors.
Such corrections are constrained by algebraic consistency conditions
similar to the Wess-Zumino consistency conditions for anomalies.
We have a complete argument for Weyl invariance up to spacetime dimension $d \le 10$,
and an argument for the standard Weyl transformation of local operators
only for operators of low dimension and spin.
There are possible `anomalous' Weyl transformations that cannot be ruled out by
algebraic consistency relations, with additional terms proportional to powers
of the Weyl tensor (for $d\geq 4$) (see \Eq{anomalousWeyltransintro}) or the Cotton tensor ($d=3$).

It is only a matter of algebra to extend these arguments to higher spacetime 
dimensions, and to operators with larger dimension and spin.
Extending to $d > 10$ is not of great interest, since we do not expect to find
any interacting fixed points in such high dimensions.
The most important question left open by this work is to understand the Weyl
transformations of operators with higher dimension and spin.
We have found some anomalous transformations that vanish in the conformally
flat case (see \Eqs{consistentOtransanomaly}
and \eq{consistentOtransanomalyd3}).
One interesting open question would be to show that there are no consistent
operator transformation anomalies in the conformally flat case.
It would also be very interesting if one could rule out the
anomalous transformations in the non-conformally flat case.
We leave these questions for future work.

\section*{Acknowledgements}
This project originated during the workshop
``Conformal Field Theories and Renormalization Group Flows in Dimensions $d > 2$,''
and MAL thanks the Galileo Galilei Institute of Theoretical Physics in Florence
for hospitality and a stimulating intellectual environment.
We thank
S. Carlip,
C.M. Chang,
A. Dymarsky,
and A. Waldron
for helpful discussions and  F. Wu for correspondence that helped us clarify some of the arguments in the paper.
This work was supported by
the Department of Energy under grant DE-FG02-91ER406746.

\newpage
\startappendices
\section*{Appendix: $T \equiv 0$ in Curved Spacetime for $6 < d \le 10$}
In this appendix, we extend the argument in \S\ref{ssec:T0} to $6 < d \le 10$.
In this case, there are no $\nabla^6$ terms in \Eq{Tcurved}
because they are forbidden by unitarity (for $d < 10$)
or are decoupled free fields (for $d = 10$).
The existence of the operators $X$ and $Y_i$ then allows the improvement
terms
\[
\eql{improveXY}
\!\!\!\!\!
\De S = \myint d^d x \gap \sqrt{g} \Bigl[ &
\xi R X
+ \xi_1 R \Box \tilde{Y}_1
+ \xi_5 R^2 \tilde{Y}_5 
+ \xi_6 R^{\mu\nu} R_{\mu\nu} \tilde{Y}_6
+ \xi_7 R^{\mu\nu\rho\si} R_{\mu\nu\rho\si} \tilde{Y}_7,
\Bigr],
\]
where $\tilde{Y}_i$ are linear combinations of the $Y_i$.
Other terms involving covariant derivatives
can be eliminated by integration by parts and the identity 
$\nabla_\mu R^{\mu\nu} =  \frac 12 \nabla^\nu R$. 
When we compute the contribution to the energy-momentum tensor from \Eq{commutatorsigma12}, 
we need to know the  metric dependence of the operators $X$ and $Y_i$.
In fact, because we are only interested in the trace $T$, it is sufficient to know
the transformation of $X$ and $Y$ under a Weyl transformation.
This question is discussed in detail in \S\ref{ssec:Weyltrans},
so we only quote the results here.
Unitarity bounds and the limit of conformal transformations in flat spacetime 
imply that under an infinitesimal Weyl transformation
$\de g_{\mu\nu} = 2 \si g_{\mu\nu}$, the most general form for
the transformation of $X$ is
\[
\de_\si X = -(d-2) \si X + \si R Y'
+ \Box \si Y'',
\]
where $Y', Y''$ are primary operators of dimension $d-4$.
Imposing commutativity of Weyl transformations, and making operator redefinitions,
one obtains the standard transformation law $\de_\si X = -(d-2) \si X$
(see the discussion below \Eq{tildedeltaO}).
%
%
Similar arguments hold for the operators $Y_i$,
and we conclude that we can compute the trace
of the energy-momentum tensor from \Eq{improveXY} assuming that the operators do not
depend on the metric.

%

The terms in \Eq{improveXY} that are linear in the curvature
will give a correction to $T$ in flat spacetime:
\[
\De T = -2(d-1) (\xi \Box X + \xi_1 \Box^2 \tilde{Y}_1) + O(R).
\]
The condition that $T \equiv 0$
in flat spacetime therefore requires $\xi,\xi_1 = 0$.
The remaining terms in \Eq{improveXY} can be used to eliminate the terms in $T$
that are quadratic in curvature, and we can simplify $T$ to 
\[
T = RX + R \Box Y_1 
+ R^{\mu\nu} \nabla_\mu \nabla_\nu Y_2
+ \nabla^\mu R \nabla_\mu Y_3 
+ \Box R Y_4.
\]
This operator relation implies the following Weyl redundancy symmetry for the sources 
for $X$ and $Y_i$:
\[
\begin{split}
\de g_{\mu\nu} &= 2\si g_{\mu\nu},
\\
\de \rho_X &= \si R,
\\
\de \rho_{Y_1} &= \si \Box R + 2 \nabla^\mu \si \gap \nabla_\mu R + R \Box \si,
\\
\de \rho_{Y_2} &= \sfrac 12 \si \Box R
+ \nabla^\mu \si \gap \nabla_\mu R + R^{\mu\nu} \nabla_\mu \nabla_\nu \si,
\\
\de \rho_{Y_3} &= -\si \Box R - \nabla^\mu \si \gap \nabla_\mu R,
\\
\de \rho_{Y_4} &= \si \Box R.
\end{split}
\]
The commutator of two such symmetries must be a symmetry, which implies that
the effective action must be invariant under the transformation
\[
\eql{rhotranscommXY}
\begin{split}
[\de_{\si_1}, \de_{\si_2}] g_{\mu\nu} &= 0,
\\
[\de_{\si_1}, \de_{\si_2}] \rho_X &= 
2(d-1) \nabla^\mu f_\mu,
\\
[\de_{\si_1}, \de_{\si_2}] \rho_{Y_1} &= 
2(d-1) \Box \nabla^\mu f_\mu 
- 2 R \nabla^\mu f_\mu 
- (d+2) \nabla^\mu R \gap f_\mu,
\\
[\de_{\si_1}, \de_{\si_2}] \rho_{Y_2} &= 
(d-1) \Box \nabla^\mu f_\mu 
+ R \nabla^\mu f_\mu
- 4 R^{\mu\nu} \nabla_\mu f_\nu
- \sfrac 12 (d+2) \nabla^\mu R \gap f_\mu,
\\
[\de_{\si_1}, \de_{\si_2}] \rho_{Y_3} &= 
-(d-1) \Box \nabla^\mu f_\mu 
- (d-1) h
- 2 R \nabla^\mu f_\mu 
+ (d-2) \nabla^\mu R \gap f_\mu,
\\
[\de_{\si_1}, \de_{\si_2}] \rho_{Y_4} &= 
2(d-1) h
+ 2 R \nabla^\mu f_\mu 
- (d-6) \nabla^\mu R \gap f_\mu,
\end{split}
\]
where we define the functions
\[
f_\mu = \si_1 \nabla_\mu \si_2 - \si_2 \nabla_\mu \si_1,
\qquad
h = \si_1 \Box^2 \si_2 - \si_2 \Box^2 \si_1.
\]
We again have a symmetry that acts only on the sources $\rho_X$ and $\rho_{Y_i}$,
but there is not sufficient freedom in choosing $\si_1$ and $\si_2$ to 
make $[\de_{\si_1}, \de_{\si_2}] \rho_i$ independent and arbitrary functions,
for a fixed metric $g_{\mu\nu}$.

One possible approach is to consider higher commutators of the symmetry,
which gives additional symmetry transformations depending on more parameters.
We will instead give an argument that is based on the fact that \Eq{rhotranscommXY} 
holds for arbitrary background metrics.
First let us consider the transformations \Eq{rhotranscommXY} in flat spacetime.
In that case, the action in the path integral transforms as
\[
[\de_{\si_1}, \de_{\si_2}]  S &= (d-1) \myint d^d x \gap \bigl[ 2\d^\mu f_\mu  (X + \Box Y)
+ h(2 Y_4-Y_3 ) \bigr],
\]
where we have integrated by parts and defined
\[
Y = Y_1 + \sfrac 12 Y_2 - \sfrac 12 Y_3,
\]
The functions $\d^\mu f_\mu$ and $h$ can be chosen independently, 
which implies the operator relations
\[
\eql{Yopreln1}
X + \Box Y \equiv 0,
\qquad
2 Y_4-Y_3  \equiv 0.
\]
Generically, this implies that we can eliminate $X$ and $Y_4$,
so that we have $Y_1, Y_2, Y_3$ as independent operators.
If $X \equiv 0$, the relation $\Box Y \equiv 0$ implies $Y \equiv 0$
and we can take $Y_1, Y_2$ as independent.
We will consider the generic case where  $Y_1, Y_2, Y_3$ are all present.
In that case, the path integral action is invariant in flat spacetime,
but in curved spacetime has variation
\[
\eql{deScurved}
[\de_{\si_1}, \de_{\si_2}]  S &= \myint d^d x \gap \sqrt{g} \ggap \biggl\{
R \nabla^\mu f_\mu \bigl[ -2Y_1 + Y_2 - Y_3 \bigr]
\nn
&\qquad\qquad\qquad{}
- 4 R^{\mu\nu} \nabla_\mu f_\nu \ggap Y_2
\nn
&\qquad\qquad\qquad{}
+ \sfrac 12 (d+2) f^\mu \nabla_\mu R \left[ - 2Y_1 - Y_2 + Y_3 \right] \biggr\}.
\]
The action is now invariant in flat spacetime, but is not invariant in a general
spacetime.
For example, we can consider the case of a maximally symmetric spacetime,
{\it i.e.\/}~Euclidean de Sitter or anti-de Sitter.
In this case, we have $R = \text{constant}$ and $R_{\mu\nu} = \frac 1d g_{\mu\nu} R$,
so we have
\[
[\de_{\si_1}, \de_{\si_2}]  S &= - \frac 1 d \myint d^d x \gap \sqrt{g} \ggap 
R \nabla^\mu f_\mu \gap \bigl[ 2d Y_1 -(d-4) Y_2 +d Y_3 \bigr].
\]
At least in a maximally symmetric space, we therefore have
the operator identity
\[
\eql{deSitterreln}
Y' \equiv 0,
\qquad 
Y' = 2d Y_1 -(d-4) Y_2 +d Y_3.
\]
But now we can take the flat limit.
All of the correlation functions of $Y'$ vanish 
identically for all nonzero values of the curvature, so
they must also vanish in flat spacetime.
We conclude that $Y' \equiv 0$ in flat spacetime.
But unitarity bounds (for $d < 10$) or decoupling (for $d = 10$)
do not allow $Y'$ to be proportional to curvature terms,
so $Y' \equiv 0$ in a general background metric.
We now have two independent operators, $Y_1$ and $Y_2$ say, with
\[
[\de_{\si_1}, \de_{\si_2}]  S &=\frac 1 d \myint d^d x \gap \sqrt{g} \ggap \biggl\{
4\nabla_\mu f_\nu \bigl(g^{\mu\nu} R- d R^{\mu\nu}  \bigr)  Y_2
-2(d+2) f^\mu \nabla_\mu R \bigl(dY_1 + Y_2\bigr) \biggr\}.
\]
It is clear that we can repeat the above argument
by considering less symmetric metrics, and conclude that
$Y_1, Y_2 \equiv 0$ for all $d \le 10$.

\newpage\frenchspacing
\bibliographystyle{utphys}
\bibliography{mycites}

\providecommand{\href}[2]{#2}\begingroup\raggedright\begin{thebibliography}{10}

\bibitem{Gross:1970tb}
D.~J. Gross and J.~Wess, ``{Scale invariance, conformal invariance, and the
  high-energy behavior of scattering amplitudes},''
\href{http://dx.doi.org/10.1103/PhysRevD.2.753}{{\em Phys. Rev.} {\bf D2}
  (1970)  753--764}.

\bibitem{Callan:1970ze}
C.~G. Callan, Jr., S.~R. Coleman, and R.~Jackiw, ``{A New improved energy -
  momentum tensor},''
\href{http://dx.doi.org/10.1016/0003-4916(70)90394-5}{{\em Annals Phys.} {\bf
  59} (1970)  42--73}.

\bibitem{Coleman:1970je}
S.~R. Coleman and R.~Jackiw, ``{Why dilatation generators do not generate
  dilatations?},''
\href{http://dx.doi.org/10.1016/0003-4916(71)90153-9}{{\em Annals Phys.} {\bf
  67} (1971)  552--598}.

\bibitem{Iorio:1996ad}
A.~Iorio, L.~O'Raifeartaigh, I.~Sachs, and C.~Wiesendanger, ``{Weyl gauging and
  conformal invariance},''
  \href{http://dx.doi.org/10.1016/S0550-3213(97)00190-9}{{\em Nucl. Phys.} {\bf
  B495} (1997)  433--450}, \href{http://arxiv.org/abs/hep-th/9607110}{{\tt
  arXiv:hep-th/9607110}}.

\bibitem{Cosme:2015cxa}
C.~Cosme, J.~M. V.~P. Lopes, and J.~Penedones, ``{Conformal symmetry of the
  critical 3D Ising model inside a sphere},''
  \href{http://dx.doi.org/10.1007/JHEP08(2015)022}{{\em JHEP} {\bf 08} (2015)
  022},
\href{http://arxiv.org/abs/1503.02011}{{\tt arXiv:1503.02011 [hep-th]}}.

\bibitem{Kos:2016ysd}
F.~Kos, D.~Poland, D.~Simmons-Duffin, and A.~Vichi, ``{Precision Islands in the
  Ising and $O(N)$ Models},''
  \href{http://dx.doi.org/10.1007/JHEP08(2016)036}{{\em JHEP} {\bf 08} (2016)
  036},
\href{http://arxiv.org/abs/1603.04436}{{\tt arXiv:1603.04436 [hep-th]}}.

\bibitem{Polchinski:1987dy}
J.~Polchinski, ``{Scale and conformal invariance in quantum field theory},''
\href{http://dx.doi.org/10.1016/0550-3213(88)90179-4}{{\em Nucl.Phys.} {\bf
  B303} (1988)  226}.

\bibitem{Luty:2012ww}
M.~A. Luty, J.~Polchinski, and R.~Rattazzi, ``{The $a$-theorem and the
  Asymptotics of 4D Quantum Field Theory},''
  \href{http://dx.doi.org/10.1007/JHEP01(2013)152}{{\em JHEP} {\bf 01} (2013)
  152}, \href{http://arxiv.org/abs/1204.5221}{{\tt arXiv:1204.5221}}.

\bibitem{Dymarsky:2013pqa}
A.~Dymarsky, Z.~Komargodski, A.~Schwimmer, and S.~Theisen, ``{On Scale and
  Conformal Invariance in Four Dimensions},''
  \href{http://dx.doi.org/10.1007/JHEP10(2015)171}{{\em JHEP} {\bf 10} (2015)
  171}, \href{http://arxiv.org/abs/1309.2921}{{\tt arXiv:1309.2921}}.

\bibitem{Dymarsky:2014zja}
A.~Dymarsky, K.~Farnsworth, Z.~Komargodski, M.~A. Luty, and V.~Prilepina,
  ``{Scale Invariance, Conformality, and Generalized Free Fields},''
  \href{http://dx.doi.org/10.1007/JHEP02(2016)099}{{\em JHEP} {\bf 02} (2016)
  099}, \href{http://arxiv.org/abs/1402.6322}{{\tt arXiv:1402.6322}}.

\bibitem{Yonekura:2014tha}
K.~Yonekura, ``{Unitarity, Locality, and Scale versus Conformal Invariance in
  Four Dimensions},''
\href{http://arxiv.org/abs/1403.4939}{{\tt arXiv:1403.4939 [hep-th]}}.

\bibitem{Jack:1990eb}
I.~Jack and H.~Osborn, ``{Analogs for the $c$ Theorem for Four-dimensional
  Renormalizable Field Theories},''
\href{http://dx.doi.org/10.1016/0550-3213(90)90584-Z}{{\em Nucl.Phys.} {\bf
  B343} (1990)  647--688}.

\bibitem{Osborn:1991gm}
H.~Osborn, ``{Weyl consistency conditions and a local renormalization group
  equation for general renormalizable field theories},''
\href{http://dx.doi.org/10.1016/0550-3213(91)80030-P}{{\em Nucl.Phys.} {\bf
  B363} (1991)  486--526}.

\bibitem{Fortin:2012hn}
J.-F. Fortin, B.~Grinstein, and A.~Stergiou, ``{Limit Cycles and Conformal
  Invariance},'' \href{http://dx.doi.org/10.1007/JHEP01(2013)184}{{\em JHEP}
  {\bf 01} (2013)  184}, \href{http://arxiv.org/abs/1208.3674}{{\tt
  arXiv:1208.3674}}.

\bibitem{Baume:2014rla}
F.~Baume, B.~Keren-Zur, R.~Rattazzi, and L.~Vitale, ``{The local
  Callan-Symanzik equation: structure and applications},''
  \href{http://dx.doi.org/10.1007/JHEP08(2014)152}{{\em JHEP} {\bf 08} (2014)
  152},
\href{http://arxiv.org/abs/1401.5983}{{\tt arXiv:1401.5983 [hep-th]}}.

\bibitem{Stergiou:2016uqq}
A.~Stergiou, D.~Stone, and L.~G. Vitale, ``{Constraints on Perturbative RG
  Flows in Six Dimensions},''
  \href{http://dx.doi.org/10.1007/JHEP08(2016)010}{{\em JHEP} {\bf 08} (2016)
  010},
\href{http://arxiv.org/abs/1604.01782}{{\tt arXiv:1604.01782 [hep-th]}}.

\bibitem{Elvang:2012st}
H.~Elvang, D.~Z. Freedman, L.-Y. Hung, M.~Kiermaier, R.~C. Myers, {\em et al.}
  \href{http://dx.doi.org/10.1007/JHEP10(2012)011}{{\em JHEP}  011},
  \href{http://arxiv.org/abs/1205.3994}{{\tt arXiv:1205.3994}}.

\bibitem{Cordova:2015fha}
C.~Cordova, T.~T. Dumitrescu, and K.~Intriligator, ``{Anomalies,
  renormalization group flows, and the a-theorem in six-dimensional (1, 0)
  theories},'' \href{http://dx.doi.org/10.1007/JHEP10(2016)080}{{\em JHEP} {\bf
  10} (2016)  080},
\href{http://arxiv.org/abs/1506.03807}{{\tt arXiv:1506.03807 [hep-th]}}.

\bibitem{Nakayama:2013is}
Y.~Nakayama, ``{Scale invariance {\it vs\/}.~conformal invariance},''
  \href{http://dx.doi.org/10.1016/j.physrep.2014.12.003}{{\em Phys. Rept.} {\bf
  569} (2015)  1--93},
\href{http://arxiv.org/abs/1302.0884}{{\tt arXiv:1302.0884 [hep-th]}}.

\bibitem{GrahamConf}
C.~Graham, ``{Conformally Invariant Powers of the Laplacian II:
  Nonexistence},'' {\em J. London Math. Soc.} {\bf s2-46 (3)} (1992)
  566Ð--576.

\bibitem{Eastwood:2002su}
M.~G. Eastwood, ``{Higher symmetries of the Laplacian},''
  \href{http://dx.doi.org/10.4007/annals.2005.161.1645}{{\em Annals Math.} {\bf
  161} (2005)  1645--1665},
\href{http://arxiv.org/abs/hep-th/0206233}{{\tt arXiv:hep-th/0206233
  [hep-th]}}.

\bibitem{GoverConf}
A.~Gover and K.~Hirachi, ``{Conformally Invariant Powers of the Laplacian Ð A
  Complete Nonexistence Theorem},'' {\em J. Am. Math. Soc.} {\bf 17 (2)} (2004)
   389Ð--405.

\bibitem{Karananas:2015ioa}
G.~K. Karananas and A.~Monin, ``{Weyl vs. Conformal},''
  \href{http://dx.doi.org/10.1016/j.physletb.2016.04.001}{{\em Phys. Lett.}
  {\bf B757} (2016)  257--260}, \href{http://arxiv.org/abs/1510.08042}{{\tt
  arXiv:1510.08042}}.

\bibitem{Brust:2016gjy}
C.~Brust and K.~Hinterbichler, ``{Free $\Box^k$ Scalar Conformal Field
  Theory},'' \href{http://arxiv.org/abs/1607.07439}{{\tt arXiv:1607.07439}}.

\bibitem{Jackiw:2005su}
R.~Jackiw, ``{Weyl symmetry and the Liouville theory},''
  \href{http://dx.doi.org/10.1007/s11232-006-0090-9}{{\em Theor. Math. Phys.}
  {\bf 148} (2006)  941--947}, \href{http://arxiv.org/abs/hep-th/0511065}{{\tt
  arXiv:hep-th/0511065}}.

\bibitem{Edery:2014nha}
A.~Edery and Y.~Nakayama, ``{Restricted Weyl invariance in four-dimensional
  curved spacetime},'' \href{http://dx.doi.org/10.1103/PhysRevD.90.043007}{{\em
  Phys. Rev.} {\bf D90} (2014)  043007},
  \href{http://arxiv.org/abs/1406.0060}{{\tt arXiv:1406.0060}}.

\bibitem{Karananas:2015eha}
G.~K. Karananas and A.~Monin, ``{Weyl and Ricci gauging from the coset
  construction},'' \href{http://dx.doi.org/10.1103/PhysRevD.93.064013}{{\em
  Phys. Rev.} {\bf D93} (2016) no.~6, 064013},
\href{http://arxiv.org/abs/1510.07589}{{\tt arXiv:1510.07589 [hep-th]}}.

\bibitem{Picco:2012ak}
M.~Picco, ``{Critical behavior of the Ising model with long range
  interactions},'' \href{http://arxiv.org/abs/1207.1018}{{\tt
  arXiv:1207.1018}}.

\bibitem{ElShowk:2011gz}
S.~El-Showk, Y.~Nakayama, and S.~Rychkov, ``{What Maxwell Theory in $D \ne 4$
  teaches us about scale and conformal invariance},''
  \href{http://dx.doi.org/10.1016/j.nuclphysb.2011.03.008}{{\em Nucl. Phys.}
  {\bf B848} (2011)  578--593}, \href{http://arxiv.org/abs/1101.5385}{{\tt
  arXiv:1101.5385}}.

\bibitem{WessCFT}
J.~Wess, ``{The conformal invariance in quantum field theory},'' {\em Nuovo
  Cim.} {\bf 18} (1960)  1086--1107.

\bibitem{Mack:1969rr}
G.~Mack and A.~Salam, ``{Finite component field representations of the
  conformal group},''
\href{http://dx.doi.org/10.1016/0003-4916(69)90278-4}{{\em Annals Phys.} {\bf
  53} (1969)  174--202}.

\bibitem{Mack:1975je}
G.~Mack, ``{All unitary ray representations of the conformal group $SU(2,2)$
  with positive energy},''
\href{http://dx.doi.org/10.1007/BF01613145}{{\em Commun.Math.Phys.} {\bf 55}
  (1977)  1}.

\bibitem{Minwalla:1997ka}
S.~Minwalla, ``{Restrictions imposed by superconformal invariance on quantum
  field theories},'' {\em Adv. Theor. Math. Phys.} {\bf 2} (1998)  781--846,
  \href{http://arxiv.org/abs/hep-th/9712074}{{\tt arXiv:hep-th/9712074}}.

\bibitem{Capper:1974ic}
D.~M. Capper and M.~J. Duff, ``{Trace anomalies in dimensional
  regularization},''
\href{http://dx.doi.org/10.1007/BF02748300}{{\em Nuovo Cim.} {\bf A23} (1974)
  173--183}.

\bibitem{Deser:1976yx}
S.~Deser, M.~J. Duff, and C.~J. Isham, ``{Nonlocal Conformal Anomalies},''
\href{http://dx.doi.org/10.1016/0550-3213(76)90480-6}{{\em Nucl. Phys.} {\bf
  B111} (1976)  45--55}.

\bibitem{Deser:1993yx}
S.~Deser and A.~Schwimmer, ``{Geometric classification of conformal anomalies
  in arbitrary dimensions},''
  \href{http://dx.doi.org/10.1016/0370-2693(93)90934-A}{{\em Phys. Lett.} {\bf
  B309} (1993)  279--284},
\href{http://arxiv.org/abs/hep-th/9302047}{{\tt arXiv:hep-th/9302047
  [hep-th]}}.

\bibitem{Wess:1971yu}
J.~Wess and B.~Zumino, ``{Consequences of anomalous Ward identities},''
\href{http://dx.doi.org/10.1016/0370-2693(71)90582-X}{{\em Phys. Lett.} {\bf
  37B} (1971)  95--97}.

\bibitem{Bardeen:1984pm}
W.~A. Bardeen and B.~Zumino, ``{Consistent and Covariant Anomalies in Gauge and
  Gravitational Theories},''
\href{http://dx.doi.org/10.1016/0550-3213(84)90322-5}{{\em Nucl. Phys.} {\bf
  B244} (1984)  421--453}.

\bibitem{Cappelli:1988vw}
A.~Cappelli and A.~Coste, ``{On the Stress Tensor of Conformal Field Theories
  in Higher Dimensions},''
\href{http://dx.doi.org/10.1016/0550-3213(89)90414-8}{{\em Nucl. Phys.} {\bf
  B314} (1989)  707--740}.

\end{thebibliography}\endgroup

\end{document}